# Observation of Interband Two-Photon Absorption Saturation in CdS Nanocrystals

**Jun He, Jun Mi, Heping Li, and Wei Ji***

*Department of Physics, National University of Singapore, 2 Science Drive 3, Singapore 117542, Republic of Singapore*

**ABSTRACT**

We report the observation of interband two-photon absorption (TPA) saturation in cadmium sulfide nanocrystals (CdS NCs) under intense femtosecond laser excitation with 1.6-eV photon energy. The observation has been compared to interband TPA saturation in bulk CdS under the same experimental conditions. By using both Z-scan technique and transient absorption measurement, the saturation intensity has been determined to be 190 $GW/cm^2$ for CdS NCs of 4-nm diameter, which shows two orders of magnitude greater than that for CdS bulk crystal. The results are in agreement with an inhomogeneously-broadened, saturated TPA model.

*Corresponding author. E-mail: phyjiwei@nus.edu.sg

Multiphoton absorption (MPA) processes were predicted theoretically in 1931.[1] Since then MPA has been observed in a wide variety of materials such as insulators, semiconductors, organic molecules and polymers, *etc.*.[2] Recently, MPA in both organic materials and semiconductor quantum dots has received great attention.[3–9] Because the quadratic (or higher-order) dependence of MPA on excitation intensity dictates the occurrence of MPA in a spatially confined region; and this unique property can be utilized to realize three-dimensional imaging based on two-photon-excited fluorescence.[9] In such applications, the saturation of two-photon absorption (TPA) should be anticipated at high excitation intensity due to the finite number of excited states, and it is an unwanted property. Unfortunately, the information on TPA saturation in quantum dots is unavailable in the literature, though there have been studies on bulk semiconductors, proteins and polymers.[10–12] In this Letter, we report for the first time the observation of interband TPA saturation in semiconductor nanocrystals. By performing both Z-scan and transient absorption measurements with ultrafast laser pulses, we have determined the saturation intensity for cadmium sulfide nanocrystals (CdS NCs) to be 190 GW/cm$^2$, which is the greatest value reported in the literature, and should be considered as a highly desirable factor in applications of quantum dots to three-dimensional imaging by two-photon-excited fluorescence, optical limiting based on TPA in quantum dots, *etc.*.

Both synthesis and sample characterization of CdS NCs used in our investigation have been reported previously.[13] A brief description of these CdS NCs is given as follows. The CdS NCs were embedded in a free-standing 130-$\mu$m-thick Nafion film, and were concentrated within two 0.9-$\mu$m-thin layers in the film: one was located close to the front surface and the other was near to the back surface of the film. The average diameter of the CdS NCs was 4.1 nm with a narrow size dispersion (half width at half maximum = 0.5 nm). The lowest interband transition for one-photon absorption was peaked at 2.71 eV, consistent with the theoretical calculation.[14] In our previous study,[13] we have observed TPA, two-photon-excited electron and hole effects, optical Kerr nonlinearity and 1-ps recovery time at excitation levels (~ 2 GW/cm$^2$).

Here, we investigate TPA saturation in the CdS NCs at room temperature with Z-scan technique[15,16] at higher excitation irradiances (up to 500 GW/cm$^2$, or 60 mJ/cm$^2$). The 1-mJ, 120-fs laser pulses were generated by a Ti:Sapphire regenerative amplifier (Quantronix, Titan), which was seeded by an erbium-doped fiber laser (Quantronix, IMRA). The laser pulses were delivered at 1 kHz pulse repetition rate and attenuated with a set of neutral density filters. Then they were focused onto the CdS-NC-embedded film. The spatial distribution of the pulses was nearly Gaussian, after passing through a spatial filter. The high quality of the Gaussian beam was confirmed by the $M^2$-factor measurement, showing that the value of $M^2$ was close to the unity. The laser beam was divided by a beam splitter into two parts. The reflected part was taken as the reference representing the incident pulse energy and the transmitted beam was focused through the CdS-NC-embedded film. Two light power probes (Laser Probe, RkP-465 HD) were used to record the incident and transmitted laser power simultaneously. A computer-controlled translation stage was employed to move the film along the propagation direction (Z-axis) of the laser pulses. For the open-aperture Z scans, extreme care was taken to ensure the entire collection of the transmitted light, and thus, self-lensing effects were eliminated. In addition, laser-induced permanent damage was studied with a Z-scan method reported in Ref 17. We found that laser-induced damage occurs at excitation irradiances (before the sample surface) of ~500 GW/cm$^2$ (or ~60 mJ/cm$^2$) or higher. All the Z scans reported here were performed with excitation irradiances below the damage threshold.

Figure 1(a) illustrates Z-scan curves measured for the CdS NCs at two excitation irradiances ($I_{00}$), where $I_{00}$ is denoted as the peak, on-axis irradiance at the focal point before the sample's front surface. Note that the Z-scan curves are imperfect; there are two background shoulders on either side of the signal, and the right one is more pronounced. This can be attributed to a poor surface quality of the film in which the CdS NCs are embedded. Such "parasitic" effects have been observed and can be eliminated by subtracting the low-irradiance background Z scan from the high-irradiance one.[15] The inset in Fig. 1(a) shows perfect Z-scan signals after subtractions. For comparison, similar measurements were conducted on a 0.5-mm-thick hexagonal CdS bulk crystal (Semiconductor Wafer, Inc.), as shown in Fig. 1(b), with laser polarization perpendicular to its *c* axis. To show the TPA saturation clearly, we plot the reciprocal energy transmittance ($T^{-1}$) as a function of the incident irradiance ($I_0$) by converting the Z-scan curves with $I_0 = \frac{I_{00}}{1+z^2/z_0^2}$ and $z_0 = \frac{\pi w_0^2}{\lambda}$, where $w_0$ =24.4 ± 0.5 $\mu$m is the minimum beam waist (half width at e$^{-2}$ maximum) at the focal point ($z$ = 0), and $\lambda$ the free-space wavelength. Figure 2 displays two examples: (a) one for the NCs and (b) the other for the bulk crystal.

Lami *et al.*[10] observed interband TPA saturation in bulk CdS and numerically modeled the observation to extract the TPA saturation intensity ($I_s$) by the use of $\alpha_2(I) = \frac{\alpha_2^0}{1+I/I_s}$, where $\alpha_2(I)$ and $\alpha_2^0$ are the intensity-dependent and low-intensity

TPA coefficient, respectively. To simulate the reciprocal energy transmittance in Fig. 2, we express the irradiance variation of the laser pulse inside the nonlinear medium with:

$$\frac{\partial I(z,r,t)}{\partial z'} = -\{\alpha_0 + \alpha_2[I(z,r,t)]I(z,r,t) + \sigma N_{e-h}\}I(z,r,t)\,, \qquad (1)$$

where $\alpha_0$ is the linear absorption coefficient, $\sigma$ the free-carrier absorption (FCA) cross section, and $N_{e\text{-}h}$ the free-carrier density , which is governed by

$$\frac{\partial N_{e-h}}{\partial t} = \frac{\alpha_2[I(z,r,t)]I^2(z,r,t)}{2\hbar\omega} - \frac{N_{e-h}}{\tau}\,, \qquad (2)$$

where $\tau$ is the effective relaxation time of the two-photon-excited free carriers, which includes the contribution from intraband relaxation, band-to-band recombination and band-to-defect relaxation in bulk crystals or band-to-defect/surface processes in NCs.

In solving Eqs. (1)–(2), we assume that the incident irradiance has a Gaussian temporal profile. (Note that the irradiance in the above equations is the one within the sample. To relate the internal irradiance to the incident irradiance just before the sample surface, one should take Fresnel reflections into account.) The transmitted irradiance is numerically calculated and then it is integrated spatially and temporally to give the transmitted energy. To check the validity of our numerical model, we fit the data for the CdS bulk crystal. From the best fits to the five Z scans recorded at $I_{00}$ from 14 to 180 GW/cm$^2$, the values of $\alpha_2^0$, $I_s$ and $\sigma$ are found to be 8.8 cm/GW, 6.2 GW/cm$^2$ and $3.0 \times 10^{-17}$ cm$^2$, respectively. In the model calculation, there are three free parameters, namely (1) $\alpha_2^0$, (2) $I_s$, and (3) $\sigma$. However, the $\alpha_2^0$ value was unambiguously determined by Z scans at low irradiances of a few GW/cm$^2$, where the TPA saturation and FCA are negligible. Our $\alpha_2^0$ value is very close to the ones revealed by other reports.[18–20]

Lami *et al.*[10] have pointed out that the FCA effect is insignificant. Indeed, in Fig. 2(b), we observe the small difference between the dotted and dashed curves calculated with and without FCA when the saturation intensity is set to infinitely large. It is obvious that the bending of the $T^{-1}$-*vs.*-$I_0$ curve is mainly dominated by the saturation of TPA. The saturation intensity is 6.9 GW/cm$^2$ when $\sigma = 0$; and 6.2 GW/cm$^2$ when $\sigma = 3.0 \times 10^{-17}$ cm$^2$. (The $\sigma$ value used is consistent with the one reported in Ref. 20.) Therefore, the saturation intensity is measured in our experiments with a high degree of certainty. This measured saturation intensity is one order of magnitude smaller than 65 GW/cm$^2$, observed at photon energy of 2 eV by Lami *et al.*.[10] This discrepancy is anticipated for the following reasons. (i) The density of states at 3.2 eV (two-photon energy used in our measurement) is less than that at 4.0 eV (used by Lami *et al.*), resulting in easier saturation by taking account of the difference in the density of states. (ii) The saturation intensity was determined by Lami *et al.* directly from an analytical solution which is based on the intensity transmittance instead of the energy transmittance. In fact, we would have 15 GW/cm$^2$ for the saturation intensity if we employed the same analytical method as Lami *et al.*.

By comparing Fig. 2(a) to Fig. 2(b), it is evident that it is harder to saturate TPA in NCs than in their bulk counterpart. Due to quantum confinement effect, electronic structures in quantum dots are much closer to atomic-like systems.[21] As the particle size is reduced, a number of nearby states available at slightly different energies in the bulk are compressed by quantum confinement into a single energy level in a quantum dot. Therefore, the density of states is increased tremendously; and the saturation process of state filling should be much more difficult to achieve in the NC than in the bulk. The photon energy of the laser pulses was 1.6 eV in our experiments, which makes the two-photon energy nearly resonant for interband TPA between the $P_h$ and $S_e$ states in CdS NCs.[14] Our previous study[13] has shown the presence of inhomogeneously broadening (half width at half maximum = 19 nm), in the photoluminescence signal caused by the band-edge radiative recombination. Our analysis confirms that the inhomogeneously broadening is mainly due to the size dispersion (~13%) of the NCs. If we assume that allowed two-photon transitions occur between one atomic-like energy level in the conduction band and the other in the valence band, the transitions can be approximately treated as a two-level system. For such a system, the saturation of TPA may be derived as $\alpha_2(I) = \frac{\alpha_2^0}{1 + I^2 / I_s^2}$.[11,12] Since the CdS NCs studied are not uniform in size, we analyze the Z-scan data on the CdS NCs in a similar fashion to the above discussion except that an inhomogeneous model: $\alpha_2(I) = \frac{\alpha_2^0}{\sqrt{1 + I^2 / I_s^2}}$ is adopted.[11] The best fit shown in Fig. 2(a) indicates that $\alpha_2^0$ = 10.7 cm/GW, $I_s$ = 190 GW/cm$^2$, and $\sigma = 2.0 \times 10^{-18}$ cm$^2$. Table 1 summarizes the nonlinear parameters used in the numerical simulation for both the NCs and the bulk crystal. The low-intensity TPA coefficient for the CdS NCs is 10.7 cm/GW, which is in good agreement with the previous pump-probe measurement of 9.5 cm/GW.[13] If we convert the TPA coefficient of 10.7 cm/GW to the TPA cross section, we will find that it is in the same order of magnitude as the results of Van Oijen *et al.*[7] and Chon *et al.*.[8] The FCA cross section is close to the value reported by Banfi *et al.*.[22]

In Fig. 2, we also numerically simulate the nonlinear transmittance for the following two scenarios: namely, (i) the dashed lines are calculated when both TPA saturation and FCA are ignored, and (ii) the dotted lines result from switching off TPA saturation only.

But, no suitable nonlinear parameters could be found to account in the both scenarios for the experimental results. The insets in Fig. 2 show the variation of the on-axis average density of excited electron-hole pairs versus incident intensity, calculated from the number of absorbed photons. Again, the experimental variation departs from the expected variation when no saturation is taken into account.

In order to investigate the phase variation of the laser pulses, closed-aperture Z scan technique was also performed on the CdS NCs. Figure 3 shows an open-aperture Z scan and an closed-aperture Z scan divided by the open-aperture Z scan. If the pure third-order nonlinear process is considered, the effective $n_2$ value ($3.0 \times 10^{-4}$ $cm^2$/GW, or $Re\chi^{(3)} = 1.8 \times 10^{-11}$ esu) should be extracted by using the standard Z-scan theory.[15] This result is smaller than that of the previous OKE measurement ($5.7 \times 10^{-4}$ $cm^2$/GW).[13] The discrepancy can be attributed to the different incident irradiances used. In the Z scan measurements reported here, a higher laser intensity of ~ 165 GW/$cm^2$ was used and therefore the free-carrier effect can not be ignored.[23] If we include this effect into the Z scan theory[16] with the $\alpha_2^0$ and $n_2$ values obtained from the Z-scan and previous OKE measurement,[13] respectively, the free-carrier refraction cross section ($\sigma_r$) can be determined to be $-2.6 \times 10^{-21}$ $cm^3$. This value is close to the results on $CdS_xSe_{1-x}$ nanoparticles measured by Bindra *et al.* at high irradiance of 566 GW/$cm^2$.[23] It is interesting to note that the optical Kerr nonlinearity ($n_2$) for the CdS NCs is positive at 780 nm but the free-carrier refraction is negative, consistent with the findings of Bindra *et al.*.[23] Note that in bulk semiconductors, free carriers (or free charged carriers) refer to electrons in the conduction band and holes in the valence band. For comparison reason, we adopt the same term for the NC case: it should be meant for excited electrons at $S_e$ states and holes at $P_h$ states.

Figure 4(a) and 4(b) show schematic diagrams of photo-dynamics for the NCs and the bulk crystal, respectively. The effective relaxation time is given by $1/\tau = 1/\tau_{intra} + 1/\tau_{inter} + 1/\tau_{trap} + ......$, where $\tau_{intra}$ is the intraband relaxation time, $\tau_{inter}$ the interband relaxation time and $\tau_{trap}$ the trapping time by surface (or defect) states. The intraband relaxation times have been reported to be of 140 fs[24] and 350 fs[10] for CdS NCs and CdS bulk crystal. The interband relaxation time and/or trapping times can be revealed by time-resolved pump-probe measurements. In our experiments, we used a cross-polarized, pump-probe configuration[13] with 780-nm, 120-fs laser pulses from the femtosecond laser system described previously. With the cross-polarized configuration, any "coherent artifact" on the transient signal was eliminated. Figure 4(c) and 4(d) illustrate the degenerate transient transmission signals ($-\Delta T$) as a function of the delay time for the NCs and the bulk crystal, respectively. For the bulk crystal, the transient signals clearly indicate two components. By using a two-exponential-component model, the best fits (no shown in Fig. 4) produce $\tau_1$ of ~ 120 fs and $\tau_2$ of > 100 ps. $\tau_1$ is the autocorrelation of the laser pulses used. The $\tau_2$ component is the interband relaxation time. As for the NCs, however, there is another recovery component with a characteristic time of ~ 1 ps. The exact nature of this slow component is unclear. But it is in agreement with the observation of Klimov *et al.*,[24] who attributed the 1-ps component to the trapping of photo-generated holes at shallow acceptor states of CdS NCs dispersed in a glass matrix. From these studies, we conclude that the effective relaxation time is dominated by the intraband relaxation time since it is the shortest. Therefore, in our simulation for TPA saturation, the effective relaxation time of 140 fs and 350 fs are used for the CdS NCs and the CdS bulk crystal, respectively. It should be emphasized that effect is insignificant in the simulation when τ is varied from 0.14 to 100 ps, since they all are greater than the pulse duration. In addition, we find the transient signals peaked at zero delay for both of the samples. The peak is mainly dominated by interband TPA processes. As the pump irradiance increases, the peak increases nonlinearly, indicating of saturation. The TPA saturation in the bulk crystal is much more pronounced than the NCs, consistent with our Z-scan results

Our observation confirms that it is much harder to saturate TPA process in NCs than in bulk crystal. The magnitude of $I_s$ is 190 GW/cm$^2$ in the CdS NCs while it is ~ 6 GW/cm$^2$ in the CdS bulk crystal. The saturation intensity in NCs can be interpreted quantitatively by an inhomogeneously broadened, saturated TPA model proposed by Kirkpatrick *et al.*:[11]

$$I_s^2 = \frac{\hbar\omega\pi\Delta\omega p(2\omega)N_0}{\tau_p\sqrt{\pi/2}(1+\frac{g_k}{g_n})\alpha_2^0}\frac{g_k}{g_n}, \tag{3}$$

where $g_k$ ($g_n$) is the electronic degeneracy of the upper (lower) state, $\tau_p$ is the half width at $e^{-1}$ maximum of the femtosecond laser pulse ($\tau_p$ = 72 fs), $N_0$ the concentration of the NCs, and $p(2\omega)$ the probability of a homogeneous class of absorbers having a central frequency of $2\omega$. The quantity $\Delta\omega$ is related to the dephasing time of the excitation, *i.e.*, the width of the homogeneous line shape. In our precious study,[25] we have found that the dephasing time $T_2$ to be 2.9 fs, resulting in $\Delta\omega = 2\pi/T_2 = 2.2 \times 10^{15}$ s$^{-1}$. The magnitude of $p(2\omega)$ is estimated to be $2.3 \times 10^{-15}$ s from the size distribution of the NCs measured by our photoluminescence experiments and transmission electron microscopy.[13] In our CdS NCs, $\alpha_2^0$ is 10.7 cm/GW and $N_0$ is estimated to be $8.4 \times 10^{18}$ cm$^{-3}$.[13] If we assume $g_k = g_n$, the saturation intensity is calculated to be ~ 130 GW/cm$^2$, which is in good agreement with our experimental finding. Note that we have made an assumption of $g_k = g_n$ in the above argument. For lower excitonic states, the values of exciton degeneracy factors ($g_k$ and $g_n$) are in the range of 2 ~ 6. If these degeneracy factors are used in the calculation, the final result should not be altered significantly.

In summary, interband TPA saturation has been observed in CdS NCs with a characteristic saturation intensity of 190 GW/cm$^2$ at 3.2 eV, nearly two orders of magnitude greater than that of its bulk counterpart. Our observation is in agreement with an inhomogeneously-broadened, saturated TPA model.

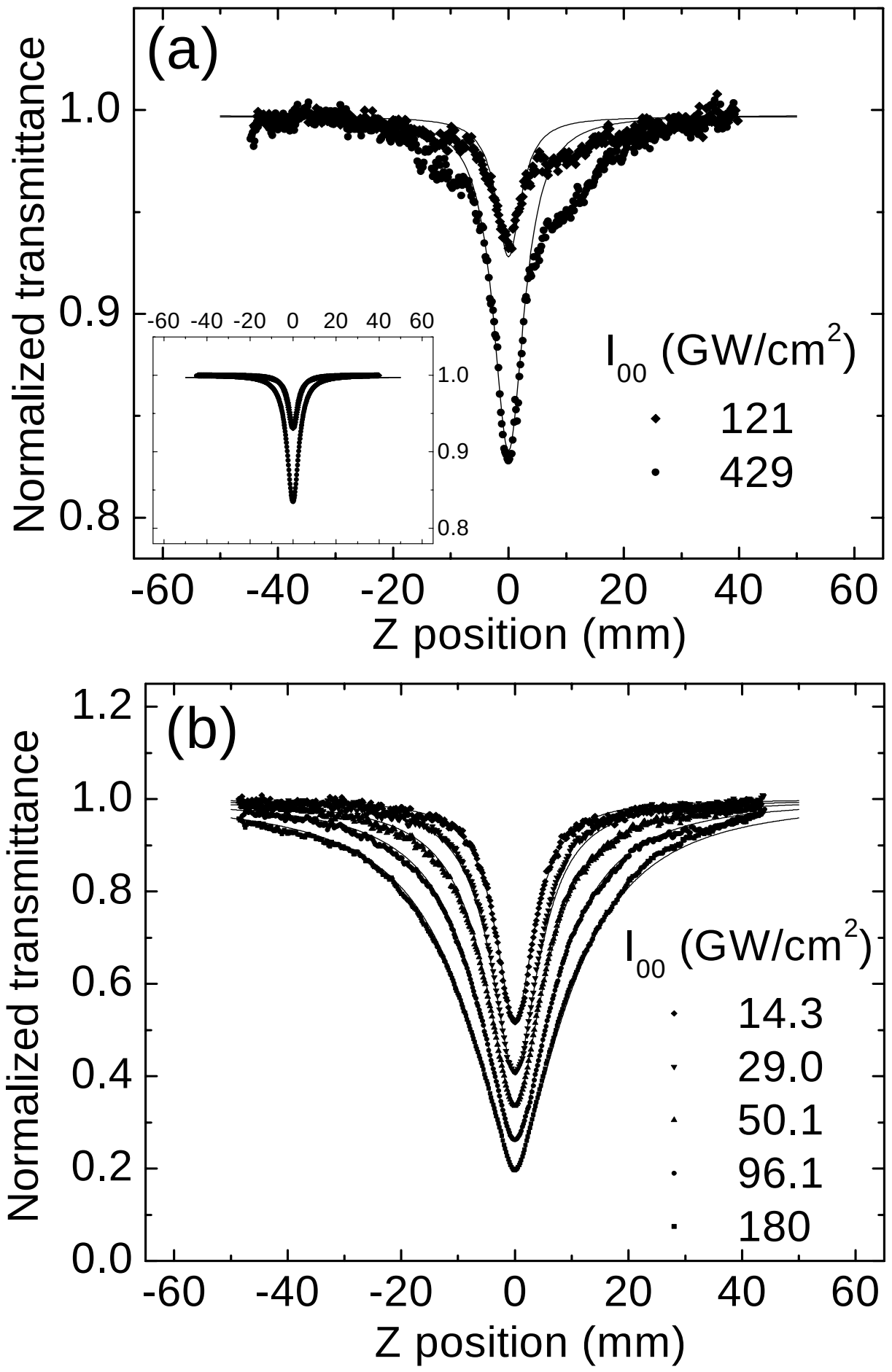

**Figure 1.** Z scans at different excitation irradiances $I_{00}$ for (a) the CdS NCs and (b) the CdS bulk crystal. The scatter graphs are experimental data while the solid lines are fitting curves calculated by Z-scan theory with saturated TPA models described in the text. The inset in (a) shows the resultant Z-scan signals after the background subtractions together with the theoretical fitting curves of the original Z-scan data.

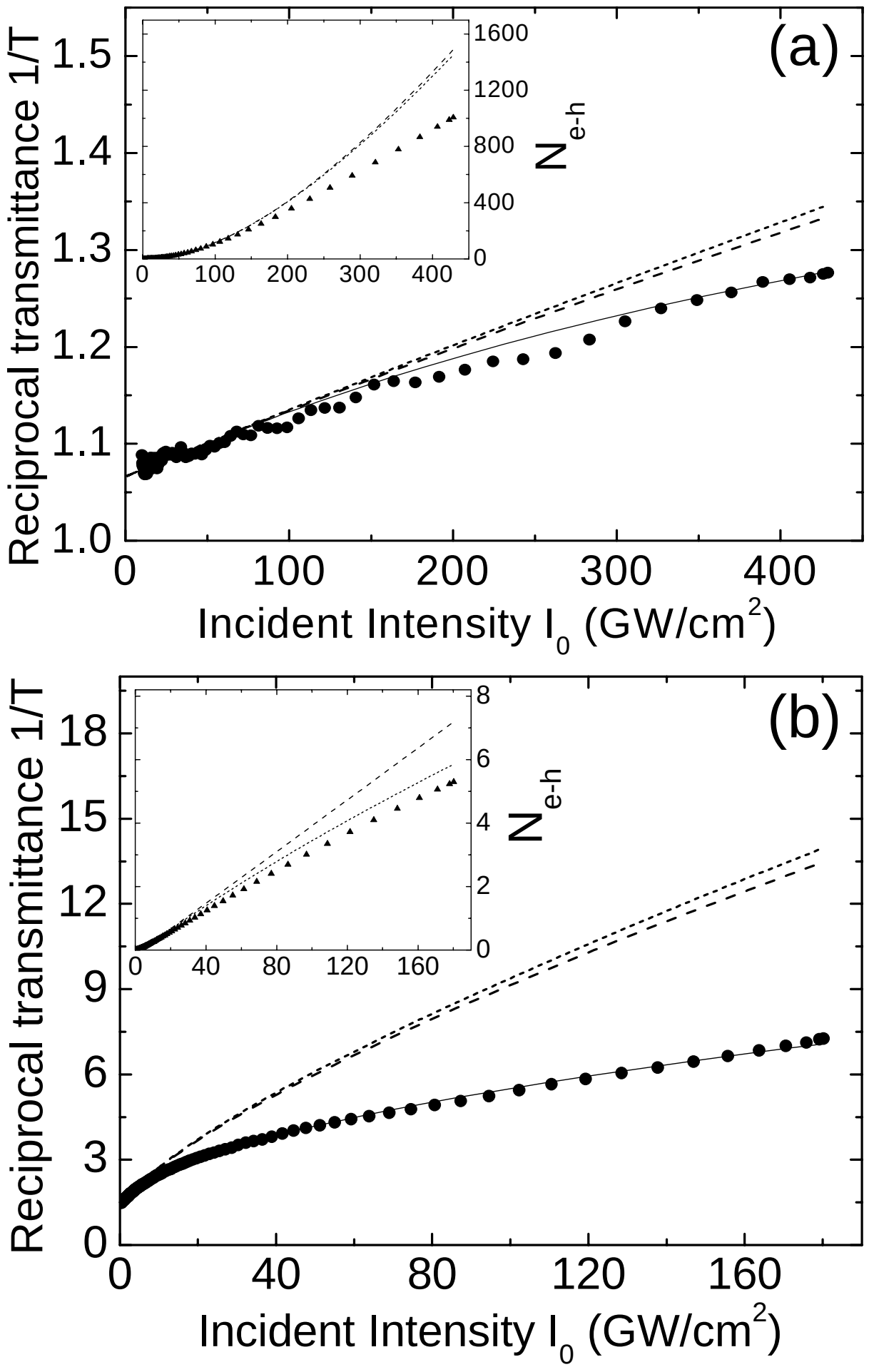

**Figure 2.** Nonlinear energy transmission and numerical modeling for (a) the CdS NCs and (b) the CdS bulk crystal. The filled circles are plots of the measured reciprocal energy transmittance ($T^{-1}$) versus the incident irradiance ($I_0$). The solid lines are theoretical fitting by the use of saturated TPA models mentioned in the text. The insets show the density $N_{e-h}$ (in $10^{17}$ cm$^{-3}$) of two-photon-created electron-hole pairs versus the incident irradiance $I_0$ (in GW/cm$^2$). The filled triangles are $N_{e-h}$ derived from the experimental data. The dashed (or dotted) lines are the theoretical variation for an unsaturated TPA without (or with) FCA.

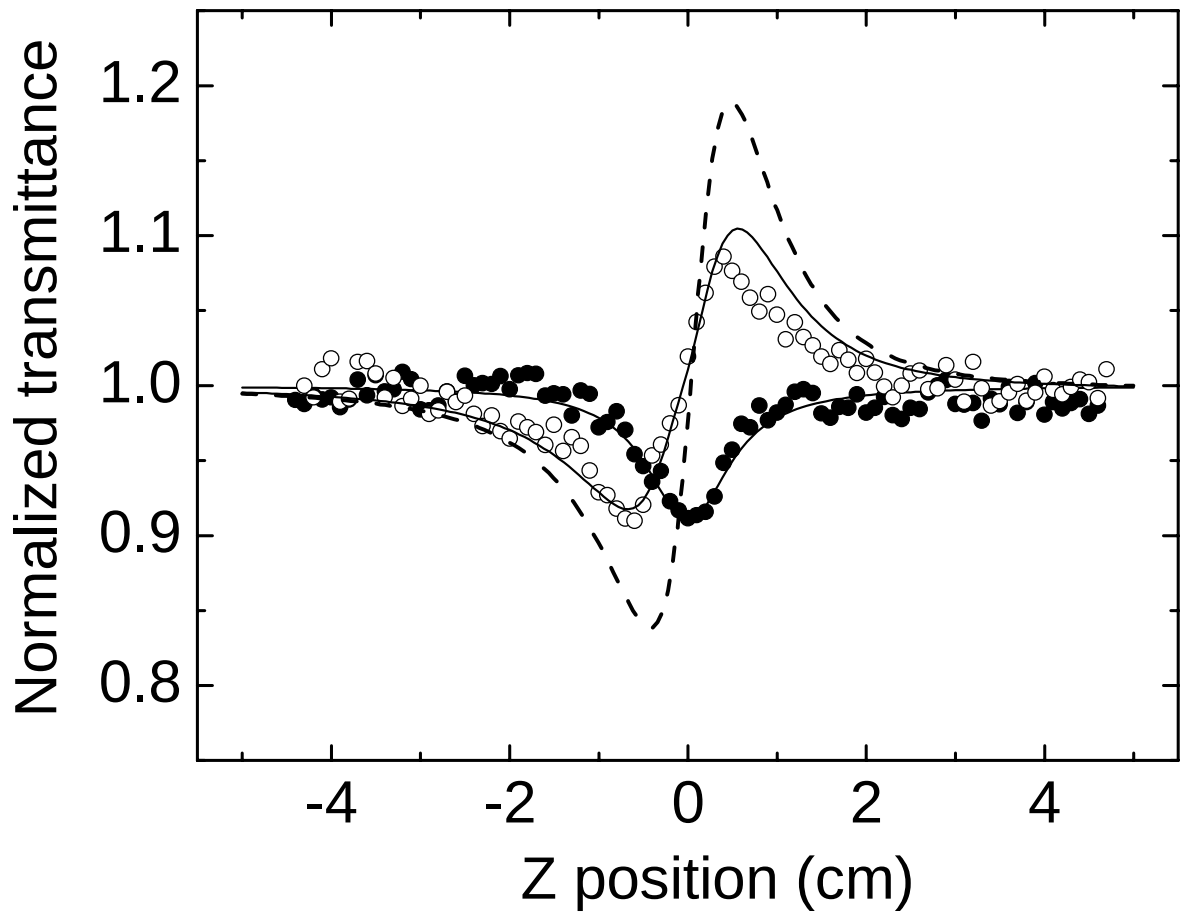

**Figure 3.** Open-aperture (filled circles) and closed-divided-open (open circles) Z-scans performed on the CdS NCs with 780-nm, 120-fs laser pulses. The excitation irradiance used is $I_{00}$ = 165 GW/cm$^2$ and the beam waist is $w_0$ = 36 $\mu$m. The solid lines are the best fits using the Z-scan theory with $\alpha_2^0$ = 10.7 cm/GW, $n_2$ = 5.7 × 10$^{-4}$ cm$^2$/GW, σ = 2.0 × 10$^{-18}$ cm$^2$ and $\sigma_r$ = –2.6 × 10$^{-21}$ cm$^3$. Dashed line is the theoretical curve obtained by neglecting the free carrier effect ($\sigma_r$ = 0 cm$^3$).

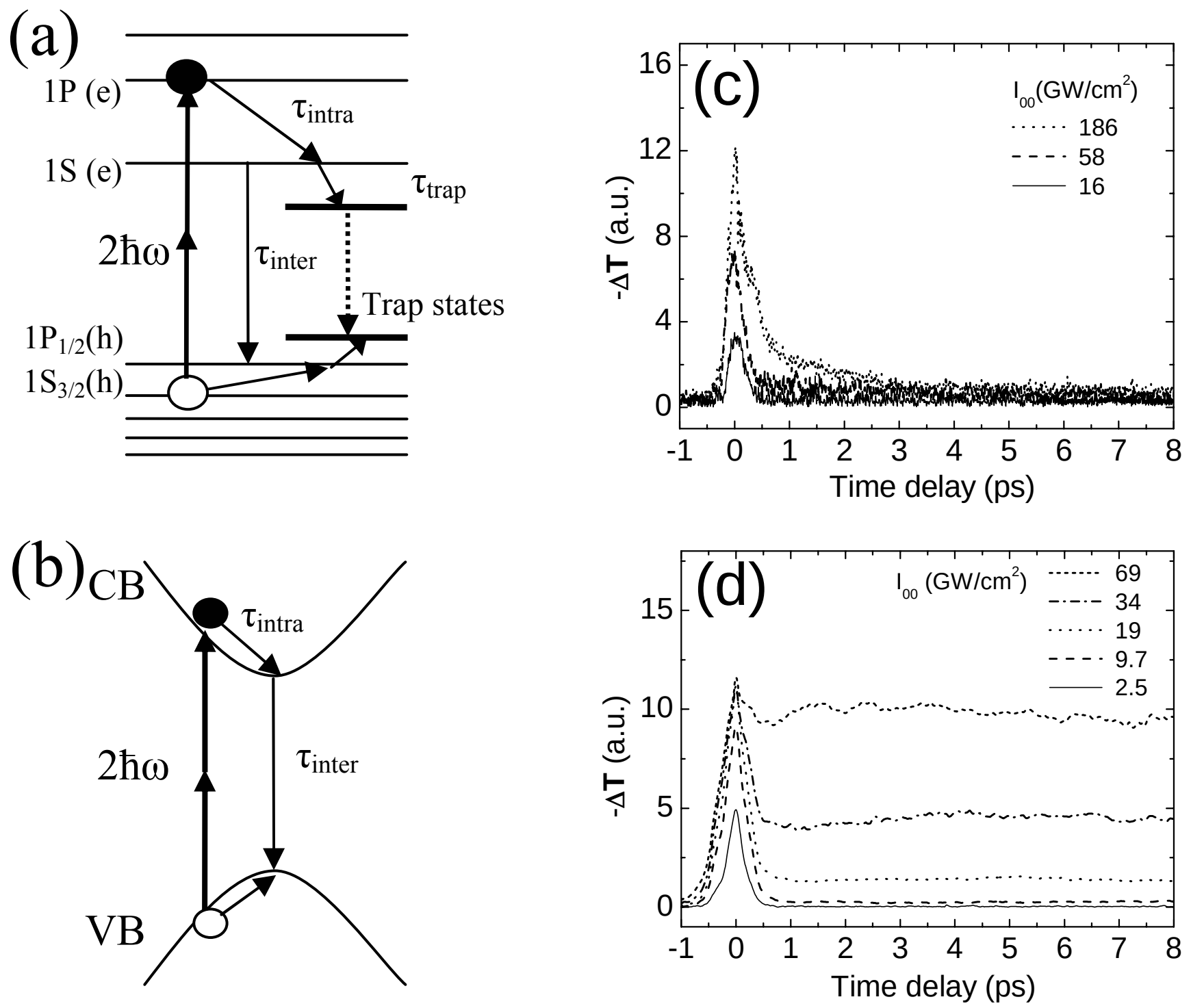

**Figure 4.** Schematic illustration of two-photon-excited carrier relaxation in (a) NCs and (b) bulk crystal. CB and VB refer to the conduction band and valence band, respectively. The experimental data in (c) and (d) are the degenerate, pump-and-probe measurements on the CdS NCs and the CdS bulk crystal, respectively, conducted at different pump irradiances ($I_{00}$).

**Table 1.** The physical parameters used in the numerical modeling with saturated TPA models described in the text.

| | Bandgap (eV) | $\tau$ (fs) | Sample thickness ($\mu$m) | $N_0$ ($cm^{-3}$) | $\alpha_2^0$ (cm/GW) | $I_s$ ($GW/cm^2$) | $\sigma$ ($10^{-18}cm^2$) | TPA cross section ($cm^4$s/photon) | $n_2$ ($10^{-5}cm^2$/GW) | $\sigma_r$ ($10^{-21}cm^3$) |
|---|---|---|---|---|---|---|---|---|---|---|
| CdS NCs | 2.71 | 140 | 1.8 | $8.4\times10^{18}$ | 10.7 | 190 | 2.0 | $7.9\times10^{-47}$ | 57 | –2.6 |
| CdS bulk | 2.42 | 350 | 500 | | 8.8 | 6.2 | 30 | | 1.7[a] | –0.8[b] |

[a]Ref. 18.

[b]Ref. 20.

## References and Notes

(1) Goppert-Mayer, M. *Ann. Phys.* (Leipzig) **1931**, *9*, 275.

(2) For example, *Handbook of Nonlinear Optics*, Sutherland, R. L. with contributions by McLean, D. G.; Kirkpatrick, S. Second Edition, Revised and Expanded (New York, NY: Marcel Dekker, 2003).

(3) Bruchez, M.; Moronne, M.; Gin, P.; Weiss, S.; Alivisatos, A. P. *Science* **1998,** *281*, 2013.

(4) He, G. S.; Markowicz, P. P.; Lin, T. C.; Prasad, P. N. *Nature* **2002**, *415*, 767.

(5) Yoshino, F.; Polyakov, S.; Liu, M.; Stegeman, G. *Phys. Rev. Lett.* **2003,** *91*, 063902.

(6) Dubertret, B.; Skourides, P.; Norris, D. J.; Noireaux, V.; Brivanlou, A. H.; Libchaber, A. *Science* **2002,** *298*, 1759.

(7) Van Oijen, A. M.; Verberk, R.; Durand, Y.; Schmidt, J.; Van Lingen, J. N. J.; Bol, A. A.; Meijerink, A. *Appl. Phys. Lett.* **2001**, *79*, 830.

(8) Chon, J. W. M.; Gu, M.; Bullen, C.; Mulvaney, P. *Appl. Phys. Lett.* **2004,** *84*, 4472.

(9) Larson, D. R.; Zipfel, W. R.; Williams, R. M.; Clark, S. W.; Bruchez, M. P.; Wise, F. W.; Webb, W. W. *Science* **2003,** *300*, 1434.

(10) Lami, J.-F.; Gilliot, P.; Hirlimann, C. *Phys. Rev. Lett.* **1996,** *77*, 1632.

(11) Kirkpatrick, S. M.; Naik, R. R.; Stone, M. O. *J. Phys. Chem. B* **2001,** *105*, 2867.

(12) Schroeder R.; Ullrich, B. *Opt. Lett.* **2002,** *27*, 1285.

(13) He, J.; Ji, W.; Ma, G. H.; Tang, S. H.; Kong, E. S. W.; Chow, S. Y.; Zhang, X. H.; Hua, Z. L.; Shi, J. L. *J. Phys. Chem. B* **2005**, *109*, 4373.

(14) Hu, Y. Z.; Koch, S. W.; Peyghambarian, N. *J. Lumin.* **1996,** *70*, 185.

(15) Sheik-Bahae, M.; Said, A. A.; Wei, T. H.; Hagan, D. J.; Van Stryland, E. W. *IEEE J. Quantum Electron.* **1990**, *26*, 760.

(16) Said, A. A.; Sheik-Bahae, M.; Hagan, D. J.; Wei, T. H.; Wang, J.; Young, J.; Van Stryland, E. W. *J. Opt. Soc. Am. B* **1992**, *9*, 405.

(17) Li, H. P.; Zhou, F.; Zhang, X. J.; Ji, W. *Opt. Commun.* **1997,** *144*, 75.

(18) Krauss, Todd D.; Wise, Frank W. *Appl. Phys. Lett.* **1994**, *65*, 1739.

(19) Van Stryland, E. W.; Woodall, M. A.; Vanherzeele, H.; Soileau, M. J. *Opt. Lett.* **1985**, *10*, 490.

(20) Li, H. P.; Kam, C. H.; Lam, Y. L.; Ji, W. *Opt. Commun.* **2001,** *190*, 351.

(21) Alivisatos, A. P. *Science* **1996**, *271,* 933.

(22) Banfi, G. P.; Degiorgio, V.; Ghigliazza, M.; Tan, H. M.; Tomaselli, A. *Phys. Rev. B* **1994**, *50*, 5699.

(23) Bindra, K. S.; Kar, A. K. *Appl. Phys. Lett.* **2001**, *79*, 3761.

(24) Klimov V.; McBranch, D. W. *Phys. Rev. B* **1997,** *55*, 13173.

(25) He, J.; Ji, W.; Ma, G. H.; Tang, S. H.; Elim, H. I.; Sun, W. X.; Zhang, Z. H.; Chin, W. S. *J. Appl. Phys.* **2004**, *95*, 6381.